\documentstyle[epsf,multicol,aps,prl,tighten]{revtex}
\renewcommand{\narrowtext}{\begin{multicols}{2}
\global\columnwidth20.5pc}
\renewcommand{\widetext}{\end{multicols}
\global\columnwidth42.5pc}  
\def\top#1{\vskip #1\begin{picture}(290,80)(80,500)\thinlines \put(   
65,500){\line( 1, 0){255}}\put(320,500){\line( 0, 1){  
5}}\end{picture}}
\def\bottom#1{\vskip #1\begin{picture}(290,80)(80,500)\thinlines \put(
330,500){\line( 1, 0){255}}\put(330,500){\line( 0, -1){
5}}\end{picture}}
\newcommand{\eq}{\begin{equation}}
\newcommand{\ee}{\end{equation}}
\newcommand{\eqa}{\begin{eqnarray}}
\newcommand{\eea}{\end{eqnarray}}
\newcommand{\s}{\sigma}
\begin{document}
\draft
\title{Compressibility of the Two-Dimensional infinite-U Hubbard
Model}
\author{Arti Tandon$^{a,b}$, Ziqiang Wang$^a$, and Gabriel Kotliar$^c$}
\address{$^a$ Department of Physics, Boston College, Chestnut Hill, MA 02167}
\address{$^b$ Department of Physics, Boston University, Boston, MA 02215}
\address{$^c$ Department of Physics, Rutgers University, Piscataway,
NJ 08854}
\date{\today}
\maketitle
\begin{abstract}
We study the interactions between the coherent quasiparticles and
the incoherent Mott-Hubbard excitations and their effects on the low
energy properties in the $U=\infty$ Hubbard model. Within the
framework of a systematic large-N expansion, these effects first
occur in the next to leading order in $1/N$. 
We calculate the scattering phase
shift and the free energy, and determine the quasiparticle
weight $Z$, mass renormalization, and the compressibility.
It is found that the compressibility is strongly renormalized and
diverges at a critical doping $\delta_c=0.07\pm0.01$.
We discuss the nature of this zero-temperature phase transition
and its connection to phase separation and superconductivity.
\end{abstract}
\narrowtext
In recent years, there has been growing interests in the physics
of doped Mott-insulators in connection with high-T$_c$ copper-oxide
superconductors. In the absence of a natural small parameter,
the relevant models of strong correlation have been extended and
studied under large symmetry groups (large N) or large dimensions
(large d). A generic feature of strong correlation is the coexistence
of coherent quasiparticles \cite{brinkmanrice}
and the broad incoherent Mott-Hubbard excitations \cite{hubbard}
that carry the main part of the spectral weight at small doping.
It has been shown in the t-J model that the systematic large-N expansion 
in the slave boson formalism provides a transparent
non-perturbative description of both the low-energy Fermi-liquid 
like quasiparticles \cite{grillikotliar} already present in the large-N limit,
and the incoherent Mott-Hubbard features at next-to-leading-order in 
1/N \cite{wangbangkotliar}.

In this paper, we study corrections to the low energy properties due
to the effects of the interactions between quasiparticles
and the incoherent Mott-Hubbard excitations by a complete calculation
of the free energy and the single particle Green's function
to next-to-leading order in 1/N.
This has not been understood properly because of the difficulty involved
in calculating the corrections to the mean-field parameters.
For simplicity, we shall consider the $U=\infty$ Hubbard model with
the spin symmetry group generalized from SU(2) to SU(N), although
the physics discussed here pertains to models that include superexchange
interactions such as the t-J model. 
This model has been solved for $N=\infty$. The ground state is a
Fermi liquid at finite hole concentrations and exhibits a Brinkman-Rice
transition at half-filling \cite{kotliarliu}. We find that the interactions
represented by the 1/N fluctuations are very strong near half-filling,
giving rise to a divergent compressibility at a finite critical doping
$\delta_c=0.07\pm0.01$ below which the Fermi liquid phase 
becomes unstable. In contrast to the Brinkman-Rice transition
at half-filling in the large-N limit, the quasiparticle residue $Z$
and the mass renormalization are only weakly renormalized and
remain finite at $\delta_c$. 
These results suggest that the Landau Fermi liquid parameters are strongly 
renormalized. In particular the instability is associated with $F_0^s\to-1$ 
as $\delta$ is reduced toward $\delta_c$, signaling the onset of 
phase-separation and/or superconductivity. 

We begin with the slave boson representation of the Hubbard model.
In the infinite-U limit, the model describes electrons with
nearest neighbor hopping, $t$, on a 2D square lattice, subject to the
constraint that double occupancy on 
each site is prohibited. It is convenient to describe the 
projected Hilbert space in terms of a neutral spin-carrying fermion,
$f_{i\s}^\dagger$, creating the singly occupied site and a spinless
charge-$e$ boson, $b_i$, keeping track of the empty site \cite{kotliarliu}.
The electron creation operator becomes $c_{i\s}^\dagger=f_{i\s}^\dagger
b_i$. In the SU(N) generalization, the occupancy constraint thus translates 
into $f_{i\s}^\dagger f_{i\s}+b_i^\dagger b_i=N/2$, where sum over
repeated $\s=1,\dots N$ index is implied. The partition function
in the coherent state path integral formulation is
\begin{equation}
{Z} = {\int}{\cal D}b^{\dagger}{\cal D}b{\cal D}f^{\dagger}{\cal D}f
{\cal D}{\lambda} e^{-\int_0^\beta L(\tau)d{\tau}},
\label{z}
\end{equation}
where the Lagrangian is given by
\begin{eqnarray}
L&=& \sum_i\left[f_{i\sigma}^{\dagger}({\partial_\tau}-{\mu})
f_{i\sigma}+b_i^\dagger\partial_\tau b_i\right]
\nonumber \\
&& 
-\frac{t}{N}\sum_{\langle i,j\rangle}\left[ f_{i\sigma}^{\dag}f_{j\sigma}
b_{j}^{\dag}b_{i} + h.c.\right]
\nonumber \\ 
&& +\sum_{i} i{\lambda}_{i}(f_{i\sigma}^{\dag}f_{i\sigma}+b_{i}^{\dag}b_{i} 
-{N/2}).
\label{l}
\end{eqnarray}	 
Here ${\lambda}_{i}$ is a static Lagrange-multiplier enforcing the 
local constraint and ${\mu}$ is the chemical potential fixing an average
of ${\delta}$ holes or $n$ particles per site, {\it i.e.}
$<f_{i\sigma}^{\dag}f_{i\sigma}> = N(1-{\delta})/2\equiv n$.
The Lagrangian in Eq.~(\ref{l}) has a U(1) gauge symmetry, it is
invariant under local U(1) transformations:
$b_i\to b_ie^{i\theta_i}$, $f_{i\s}\to f_{i\s}e^{i\theta_i}$, and
$\lambda_i\to\lambda_i-\partial_\tau\theta_i$. We choose the radial gauge
\cite{read} where the boson fields ($b_i,b_i^\dagger$) 
are replaced by a real amplitude field $r_i$ while $\lambda_i$ is promoted 
to a dynamical field $\lambda_i(\tau)$.
In this gauge, the fermionic excitations can be identified with the
Fermi liquid quasiparticles.

To enable an 1/N-expansion to the next-to-leading order, we write the
boson fields in terms of static mean-field and dynamic fluctuating parts,
\begin{equation}
r_{i}(\tau) = b[1+{\delta}r_{i}(\tau)], \quad i\lambda_{i}(\tau)= 
\lambda + i\delta\lambda_{i}(\tau).
\label{field}
\end{equation} 
In the first part of the paper, we shall calculate $b$, $\lambda$,
together with the chemical potential $\mu$ to the next-to-leading
order. 
Using these results, we then analyze the 
single-particle Green's function, determine the wave function
renormalization $Z$ and the quasiparticle mass renormalization
and the compressibility.

Substituting Eq.~(\ref{field}) into Eq.~(\ref{l}), integrating
out the fermions and the boson fields 
($\delta r,\delta\lambda$) to quadratic order in Eq.~(\ref{z}),
we obtain the free energy $F=-kT\ln Z$ to next-to-leading order
in $1/N$,
\begin{equation} 
F= -\frac{N}{{\beta}} \sum_{k,{\omega}_{n}} \ln(\epsilon_{k} -i\omega_{n}) 
+ \lambda(b^{2} - \frac{N}{2}) +F_{\rm bos},
\label{f}
\end{equation} 
where $\omega_n$ is a fermion Matsubara frequency,
${\epsilon}_{k} = -\frac{2tb^{2}}{N}{\gamma}_{k} +{\lambda} -{\mu}$ with
${\gamma}_{k}=\cos k_{x}+\cos k_{y}$, and $F_{\rm bos}$ is the contribution
due to boson fluctuations. The latter can be written in terms of the
determinant of the inverse boson propagator matrix $D^{-1}$,
\begin{equation} 
F_{\rm bos} =\frac{1}{2{\beta}} \sum_{q,{\nu_n}} \ln {\rm Det} 
D^{-1}(q,i{\nu}_{n}),
\label{fboson}
\end{equation}  
where $\nu_n$ is a boson Matsubara frequency. Note that in order to
properly regularize the theory in the radial gauge,
${\rm Det} D^{-1}$ should be evaluated on a discretized imaginary
time mesh before taking the continuum limit in $\tau$ 
\cite{longpaper,arrigoni}.
The opposite sequence of operations will lead to unphysical
ultraviolet singularities. We find,
\begin{eqnarray} 
{\rm Det} D^{-1}(q,i{\nu}_{n})&=&
P_{\lambda\lambda}(q,i\nu_{n})P_{rr}(q,i\nu_{n}) - P_{\lambda r}^2
(q,i\nu_{n})
\nonumber \\ 
&+&2b^2[\lambda-\epsilon_b(0)]S^-P_{\lambda r}(q,i\nu_n)/i\nu_n.
\label{det}
\end{eqnarray} 
Here $S^{-} = e^{-i\nu_{n}0^{-}} -  e^{i\nu_{n}0^{-}}$ 
is a regularization factor and $\epsilon_b(q)=
\lambda - 2t\sum\gamma_{k-q}n_{f}(\epsilon_{k})$ with
$n_f(\epsilon)$ the Fermi distribution function.
$P_{\alpha\beta}=N( \Pi_{\alpha\beta} + B_{\alpha\beta})$ are
the fermion polarizations given by
$B_{rr}=2b^{2}{\epsilon}_{b}(q)/N$,
$B_{\lambda r} = B_{r \lambda} = 2b^{2}/{N}$, $B_{\lambda \lambda} = 0$,
and
\eq
{\Pi}_{\alpha \beta} = \sum_{k} 
\frac{ n_f({\epsilon_{k_+}}) -  n_f({\epsilon_{k_-}})}{{\epsilon}_{k_+} - 
{\epsilon}_{k_-} - i{\nu}_{n}} {\Lambda}^{\alpha}(k,q) {\Lambda}^{\beta}(k,q),
\end{equation}
where $k_{\pm}=k\pm q/2$ and
$\Lambda=[-2tb^2/N(\gamma_{k_+}+\gamma_{k_-}),i]$ are the boson-fermion
vertices.

The values of the parameters $(b,\lambda,\mu)$ are determined by minimizing
the free energy in Eq.~(\ref{f}), leading to three self-consistent equations:
\eq
{\partial F\over\partial b}=0,\quad
{\partial F\over\partial \lambda}=0,\quad
{\partial F\over\partial \mu}=-n.
\label{mf}
\ee
Solving these equations to leading order in 1/N, where only the fermion
contribution enters Eq.~(\ref{f}), one recovers the results
of Kotliar and Liu \cite{kotliarliu}, namely,
a boson condensate $b^2=b_0^2=N\delta/2$ 
and a chemical potential shift $\lambda=\lambda_0
=2t\sum_k\gamma_k n_f(\epsilon_k)$.
This corresponds to a Fermi liquid phase with a quasiparticle
dispersion $\epsilon_k^0=-(2tb_0^2/N)\gamma_k+\lambda_0-\mu_0$
and a quasiparticle residue $Z=b_0^2=N\delta/2=m/m^*$.
The compressibility $\kappa_0=dn/d\mu=N\rho/(1+4t\rho\vert\epsilon_0\vert)$, 
where $\rho = \sum_k\delta(\epsilon_{k}^0)$ and
$\rho\epsilon_{0}=-\sum_k\gamma_{k}\delta(\epsilon_{k}^0)$.
It diverges as $\delta\to0$, together 
with $Z\to0$ and $m^*\to\infty$, giving rise to a
Brinkman-Rice \cite{brinkmanrice} metal-insulator transition.
\begin{figure}     
\vspace{-0.5truecm} 
\center     
\centerline{\epsfysize=2.6in     
\epsfbox{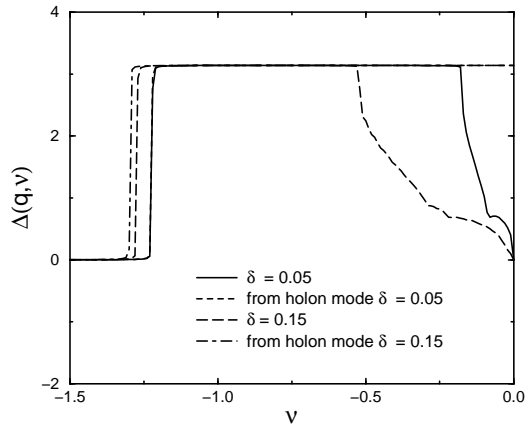}} 
\vspace{-0.5truecm}     
\begin{minipage}[t]{8.1cm}     
\caption{  
Phase shift $\Delta(q,\nu)$ at $q=(2\pi/3,2\pi/3)$ for $\delta=0.05,0.15$ and
comparison to holon contributions.}
\label{fig1}    
\end{minipage}     
\end{figure}     

The effects of interactions between the quasiparticles and the incoherent
Mott-Hubbard excitations 
enter through $F_{\rm bos}$ in Eq.~(\ref{f})
at the next-to-leading order in 1/N \cite{wangbangkotliar,wang}. 
It is instructive to rewrite
$F_{\rm bos}$ in Eq.~(\ref{fboson}) by converting the boson Matsubara
sum into a contour integral distorted along the real axis,
\begin{equation}
F_{\rm bos} =-\frac{1}{2{\pi}} \sum_{q}  
\int_{-\infty}^{\infty} d{\nu}{\Delta}(q,\nu)n_{b}(\nu),
\label{fboson2}
\end{equation}
where $n_b$ is the Bose distribution function and
\begin{equation}
{\Delta}(q,\nu)=-\arctan\left[{{\rm Im} {\rm Det}D^{-1}(q,\nu) \over {\rm Re} 
{\rm Det}D^{-1}(q,\nu)}\right]
\label{phaseshift}
\end{equation}
can be considered as a many-body phase shift
due to scattering of the fermions by particle-hole excitations.
We have numerically calculated the phase shift $\Delta$ at $T=0$ from
Eqs.~(\ref{phaseshift}) and (\ref{det}). Its general behavior is shown
in Fig.~1 for a fixed wave vector $q=(2\pi/3,2\pi/3)$ as a function
of frequency at different dopings. 
From intermediate to high frequencies,
the scattering is in the unitary limit with $\Delta=\pi$, indicating
the existence of a collective mode which is pulled out of the
particle-hole continuum at low frequency where $\Delta$ drops
from $\pi$ to zero. 
Indeed, we find that ${\rm Det} D^{-1}$ has
a branch-cut along the real axis corresponding to the particle-hole 
continuum, and isolated poles corresponding to a collective mode
which is well described by,
\begin{equation} 
\omega_q^2\simeq c^{2}[\sin^{2}(q_{x}/2) + \sin^{2}(q_{y}/2)] + 
{\epsilon}_{b}^2(q),
\label{mode}
\end{equation} 
where $c\propto \delta t$ is the zero sound velocity and $\epsilon_b(q)$
coincides with the original slave-boson dispersion.
This mode has been identified as the ``holon'' 
in the t-J model \cite{wangbangkotliar,wang}.
At small doping, the holon contribution,
with $\omega_q^*\simeq\pm\epsilon_b(q)$, dominates the particle-hole
scattering as seen in Fig.~1.
It disperses over the entire lower Hubbard band and carries the incoherent
Mott-Hubbard spectral weight. Remarkably, the holon contribution
leads to a density-density correlation function in excellent
agreement with that obtained from exact diagonalization of the
t-J model on small clusters \cite{horsch}.

Now we solve the self-consistent equations in (\ref{mf}) to 
next-to-leading order in 1/N including $F_{\rm bos}$. Writing
$b=b_{0} + b_{1}$, $\lambda = \lambda_{0} + \lambda_{1}$,
and $\mu = \mu_{0}+ \mu_{1}$, we find,
\vspace{-0.7cm}
\eqa
b_{1} &=& \frac{b_0}{2\beta}\sum_{q,{\nu}_{n}}D_{rr}(q,i{\nu}_{n}) 
e^{-i{\nu}_{n}0^{-}}
\label{b1} \\
\mu_1 &=& \lambda_1+\frac{1}{\rho}  
\sum_{k}{\Sigma}_{n}(k,{\epsilon}_{k}) 
{\delta}({\epsilon}_{k})+ \frac{4tb_{0}b_{1}{\epsilon}_{0}}{N},
\label{mu1}
\eea
\widetext
\top{-2.8cm}
\eqa
{\lambda}_{1} = &-&\frac{N}{2b_{0}^{2}{\beta}}\sum_{k,{\omega}_{n}} 
G_{0}(k,i{\omega}_{n})\bigl[{\Sigma}_{n}(k,i{\omega}_{n})
-\frac{2tb_{0}^{2}}{N} \frac{1}{\beta}\sum_{q,{\nu}_{n}}  
{\gamma}_{k-q}D_{rr}(q,i{\nu}_{n})e^{-i{\nu}_{n}0^{-}}\bigr]
\nonumber \\
&+&\frac{2t}{\beta}\sum_{k,{\omega_n}}{\gamma}_{k} 
G_{0}^2(k,i{\omega}_{n}) {\Sigma}_{n}(k,i{\omega}_{n})    
+ 2t\vert{\epsilon}_{0}\vert\sum_{k}{\Sigma}_{n}(k,{\epsilon}_{k}){\delta} 
({\epsilon}_{k}) +\frac{2}{\beta}\sum_{q,{\nu}_{n}}\left[{D}_{r \lambda} 
(q,i{\nu}_{n})-D_{r\lambda}(q,\infty)\right].
\label{lambda1}
\eea
Here $G_0^{-1}=i\omega_n-\epsilon_k^0$ and ${\Sigma}_{n}(k,i{\omega}_{n})$
is the usual self-energy to leading order in 1/N \cite{wangbangkotliar,wang},
\eqa
{\Sigma}_{n}(k,i{\omega}_{n})=&&
\frac{2tb_{0}^{2}}{N} \frac{1}{\beta}\sum_{q,i{\nu}_{n}}  
{\gamma}_{k-q}D_{rr}(q,i{\nu}_{n})e^{-i{\nu}_{n}0^{-}}  
-\frac{1}{\beta} \sum_{k,{\nu}_{n}} 
G_{0}(k+q,i{\omega}_{n}+i{\nu}_{n})\bigl[D_{\lambda \lambda} 
(q,i{\nu}_{n})S_{\lambda\lambda} 
\nonumber \\
&&+ 2D_{\lambda r}(q,i{\nu}_{n}) 
S_{r\lambda}(E_{k}+E_{k+q})   
+ D_{rr}(q,i{\nu}_{n})(E_{k}+E_{k+q})^{2}\bigr],
\label{sigman}
\eea
\bottom{-2.7cm}
\narrowtext
\noindent where $E_k=-(2tb_0^2/N)\gamma_k$,
$S_{r\lambda} = e^{-i\nu_{n}0^{-}}$, and  
$S_{\lambda\lambda} = (e^{-i\nu_{n}0^{-}} +e^{i\nu_{n}0^{-}})/2$ are
regularization factors for $D_{r\lambda}$ and $D_{\lambda\lambda}$ 
respectively. Without them, the theory in the radial gauge would be singular in
the ultraviolet because  $D_{r\lambda}$ and $D_{\lambda\lambda}$
approach constants at large frequencies \cite{longpaper}.
\begin{figure}     
\vspace{-0.5truecm} 
\center     
\centerline{\epsfysize=2.6in     
\epsfbox{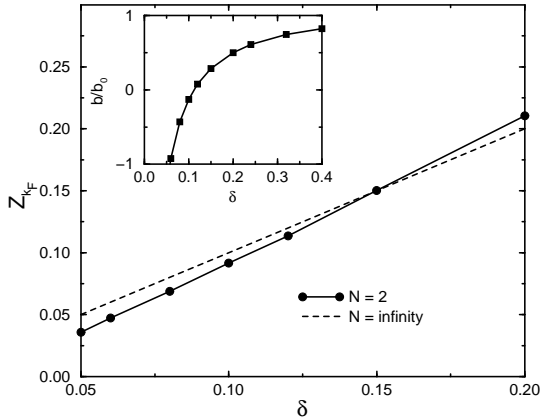}} 
\vspace{-0.5truecm}     
\begin{minipage}[t]{8.1cm}     
\caption{ Quasiparticle residue $Z_{k_{F}}$ 
as a function of doping $\delta$ in the $\Gamma$M direction. 
Inset: Slave boson condensate amplitude $b/b_0$ as a function of doping.
}   
\label{fig2}    
\end{minipage}     
\end{figure}     
Next we present the results of our numerical evaluations of 
Eqs.(\ref{b1}-\ref{lambda1}), which were done on a 2D mesh
of up to $60\times60$ points in the first quadrant of the Brillouin-zone
using the micro-zone method and a frequency grid size as small as
$\Delta\omega/t=\delta/20$ to ensure convergence.
The result for the slave-boson condensate to next-to-leading order
in 1/N is shown in the inset of Fig.~2
for $N=2$. Interestingly, $b$ vanishes at a doping 
$\delta^*\simeq0.12$. If we approximate the $D_{rr}$ in
Eq.~(\ref{b1}) by the single holon mode in Eq.~(\ref{mode}) at
small doping, we find an analytical estimate $b/b_0=1-1/4N\delta$,
which vanishes at a $\delta^*=1/4N=0.125$, in good agreement with
the numerical result. 

It is important to note that at this order,
the boson condensate is not simply related to the quasiparticle
residue, in contrast to the case in the large-N limit. To determine
the Fermi liquid coherence factor $Z$, we follow
Refs.~\cite{wangbangkotliar,wang} and write done the 1/N-resummation of
the single-electron Green's function,
\begin{equation} 
G(k,i\omega_{n}) =
\frac{b^2[1+ {\Sigma}_{a}(k,i\omega_{n})]^2}{i\omega_{n} - {\epsilon}_{k} - 
{\Sigma}_{n}(k,i{\omega}_{n})}+b^{2}{\Sigma}_{i}(k,i{\omega}_{n}),
\label{g}
\end{equation} 
where ${\Sigma}_{n}$ is given in Eq.~(\ref{sigman}),
${\Sigma}_{a}$ and ${\Sigma}_{i}$ are
the anomalous part due to the boson condensate, and the incoherent
part of the self-energies respectively. The latter are given by,
to leading order in 1/N,
\eqa
{\Sigma}_{i}(k,i\omega)&=&-T\sum_{q,\nu_{n}} G_0(k+q,i\omega +  
i\nu_{n}) D_{rr}(q,i\nu_{n}) 
\label{sigmai} \\
{\Sigma}_{a}(k,i\omega) &=& -T\sum_{q,i\nu_{n}}G_0 
(k+q,i\omega + i\nu_{n})\bigl[D_{\lambda r}(q,i\nu_{n})S_{r\lambda}
\nonumber \\
&&+( E_{k} + E_{k+q} )D_{rr}(q,i\nu_{n}) \bigr].
\label{sigmaa}
\eea
The quasiparticle residue on the interacting Fermi surface
can be obtained from Eq.~(\ref{g}),
\eq
Z_{k_{F}} = \frac{b^2[1 +
{\rm Re}{\Sigma}_{a}(k_{F},0)]^2}{\left[1 -  
{\partial {\rm Re}{\Sigma}_{n}(k_{F},\omega)}/{\partial\omega}|_{\omega \
= 0}\right]}.
\label{flz}
\ee
Thus $Z_{k_F}$ can be finite even if $b^2$ is vanishing, provided that
the reduction of the condensate is compensated by the contributions
from the self-energies. Remarkably, this turns out to be the route
followed by the 1/N-expansion. Expanding Eq.~(\ref{flz})
to next-to-leading order in 1/N, one has
\eq
Z_{k_{F}}^{1/N}=b^2+2b_0^2\Sigma_a(k_F,0)+b_0^2\partial\Sigma_n(k_F,\omega)
/\partial\omega\vert_{\omega=0}.
\label{z1overn}
\ee
In Fig.~2, $Z_{k_{F}}^{1/N}$ is plotted as a function of doping
in the $\Gamma$M direction. The 1/N-corrections are 
clearly small and $Z_{k_F}$ stays close to the large-N limit value.
Within the single holon mode (Eq.~\ref{mode}) approximation, we found
that the $1/\delta$-correction to $b^2$ in Eq.~(\ref{z1overn}) is
canceled out by the contributions from the self-energy terms, leaving
$Z_{k_{F}}^{1/N}$ weakly renormalized near $\delta^*$.
Thus we conclude that while the boson condensate vanishes at $\delta^*$, the
Fermi liquid coherence remains finite.
\begin{figure}     
\vspace{-0.5truecm} 
\center     
\centerline{\epsfysize=2.6in     
\epsfbox{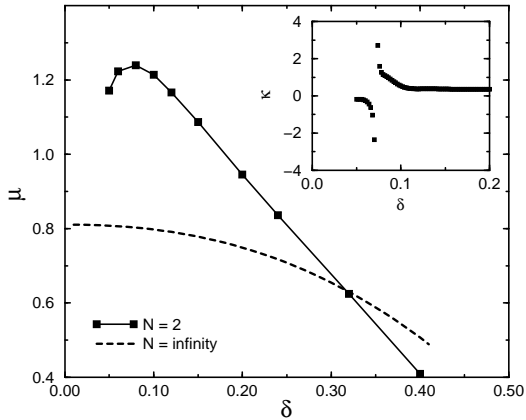}} 
\vspace{-0.5truecm}     
\begin{minipage}[t]{8.1cm}     
\caption{  
Electron chemical potential and the compressibility (inset) as a function 
of doping.}   
\label{fig3}    
\end{minipage}     
\end{figure}     

We next turn to the compressibility of the model. In Fig.~3, the electron
chemical potential $\mu=\mu_0+\mu_1$ 
is shown as a function of doping, which is strongly
modified from the $N=\infty$ result. The corresponding
compressibility $\kappa=-d\delta/d\mu$ is shown in the inset of Fig.~3. 
At moderate dopings, $\kappa$ is approximately constant, but becomes
strongly doping dependent as $\delta$ is reduced. Interestingly,
there exists a critical doping, $\delta_c=0.07\pm0.01$, 
at which $\kappa$ diverges and becomes negative for $\delta<\delta_c$.
Thus, the Fermi liquid state becomes unstable below
$\delta_c$, while no singularity is present in $Z_{k_F}$
To further understand the nature of the instability, we have 
studied the quasiparticle mass renormalization defined by
$m^*/m=N^*(0)/N(0)$, where $N(0)=\rho$ and $N^*(0)$ are the bare ($N=\infty$)
and the renormalized (next-to-leading order in 1/N) quasiparticle
density of states respectively. The numerical calculations of
$N^*(0)$ show that while $m^*$ is enhanced in the doping range
$0.05<\delta<0.2$, it does not exhibit any singular behavior. 
A well-behaved $N^*(0)$, together with the general Fermi-liquid result,
\begin{equation} 
\kappa \equiv \frac{\partial n}{\partial \mu} =   
\frac{N^{*}(0)}{1 + F^{0}_{s}},
\label{kappa}
\end{equation} 
suggests that the divergence of $\kappa$ is 
a result of the Landau Fermi liquid parameter $F_s^0\to-1$ at $\delta_c$,
indicative of phase-separation and/or superconducting instability \cite{note}.
Note that the phase separation in the infinite-U case 
has a different origin than in models with strong antiferromagnetic
correlations. For one-hole, the ground state is known rigorously to be
a Nagaoka state \cite{nagaoka} of a saturated ferromagnet.
For a finite density of holes, one expects ferromagnetic correlations
to compete with the kinetic energy and whether the Nagaoka state remains
stable is a question of great interest. 
Both numerical \cite{putikka} and analytical \cite{shastry}
results have shown that the {\it uniform} Nagaoka ferromagnetic state 
is unstable for any finite hole concentration. Our results naturally suggest 
a novel possibility that at low-doping the system phase separates into 
hole-poor ferromagnetic and hole-rich paramagnetic regions.
In the presence of long-range Coulomb repulsion, we expect the p-wave pairing
instability \cite{kotliarliu} 
enhanced by the tendency towards phase-separation to
dominate \cite{emerykivelson,grilli}.
We conclude that the breakdown of the Fermi liquid in our case is {\it not}
due to a gradual reduction of the Fermi liquid coherence, but rather the
enhanced interactions between the quasiparticles. This is the kind
of Fermi liquid instability originally envisioned by Landau.

We thank Jan Engelbrecht and Yong Baek Kim for useful discussions. 
This work was supported in part by DOE DE-FG02-99ER45747, an award from 
Research Corporation, and NSF DMR 95-29138.

\end{multicols}

\begin{references}
\bibitem{note} The actual boundary for phase separation is likely to
occur before the divergence of the compressibility after the
Maxwell construction.
\end{references}
\end{document}